\documentclass[a4paper]{article}
\usepackage{authblk}
\usepackage{amsmath}
\usepackage{graphicx}
\usepackage{natbib}
\usepackage[a4paper, left={2cm}, right={2cm}, top={2cm}, bottom={3cm}]{geometry}

\begin{document}
    \title{Analysing Meteoroid Flights Using Particle Filters}
    \author[1]{E. K. Sansom}
    \author[2]{M. G. Rutten}
    \author[1]{P. A. Bland}

    \affil[1]{Applied Geology, Curtin University, Western Australia}
    \affil[2]{Defence Science and Technology Group, Edinburgh, South Australia}
    \date{}
	\maketitle
	
    \abstract{Fireball observations from camera networks provide position and time information along the trajectory of a meteoroid that is transiting our atmosphere. The complete dynamical state of the meteoroid at each measured time can be estimated using Bayesian filtering techniques. A particle filter is a novel approach to modelling the uncertainty in meteoroid trajectories and incorporates errors in initial parameters, the dynamical model used and observed position measurements. Unlike other stochastic approaches, a particle filter does not require predefined values for initial conditions or unobservable trajectory parameters. The Bunburra Rockhole fireball \citep{Spurny2012}, observed by the Australian Desert Fireball Network (DFN) in 2007, is used to determine the effectiveness of a particle filter for use in fireball trajectory modelling. The final mass is determined to be $2.16\pm1.33\, kg$ with a final velocity of  $6030\pm216\, m\,s^{-1}$, similar to previously calculated values. The full automatability of this approach will allow an unbiased evaluation of all events observed by the DFN and lead to a better understanding of the dynamical state and size frequency distribution of asteroid and cometary debris in the inner solar system.}
	
    \section{Introduction} 

         A meteoroid is a small object moving in interplanetary space. When one enters the Earth's atmosphere, it creates a bright phenomenon called a meteor, fireball or bolide (depending on brightness). The interaction of this material with our atmosphere provides us with an opportunity to observe and study a portion of interplanetary material that would otherwise be inaccessible to us. Telescopes cannot image mm-m sized objects, and discoveries of 10’s m sized asteroids constitute a tiny fraction of the predicted population \citep{Harris2012}. Determining the physical state of this material in our atmosphere – its strength and mass distribution, and its velocity frequency distribution, provides a unique window on cometary and asteroidal material in the inner solar system. In order to derive that data, we need to model the meteoroid-atmosphere interaction. 
        
        A set of idealised equations govern how a single meteoroid body will respond in terms of velocity and mass loss. The amount of deceleration experienced by a meteoroid is related to its shape and bulk density via a \emph{shape-density parameter}, $\kappa= \frac{c_d A}{2 \rho_m^{2/3}}$, where $c_d$ is the aerodynamic drag coefficient 
        \footnote{\label{ft1}$\Gamma$ is referred to as the drag factor in many meteoroid trajectory works, including \citep{Ceplecha2005}. The aerodynamic drag coefficient, $c_d = 2\Gamma$ \citep{Bronshten1983, Borovivcka2015ast}. },
        $A$ the shape parameter as described by \citet{Bronshten1983} and the bulk density of the meteoroid- $\rho_m$. Both ablation and gross fragmentation of the meteoroid is responsible for loss of mass. Gross fragmentation is hard to predict and is linked to the strength of the object. Ablation can be quantified through the \emph{ablation parameter} - $\sigma$, which is defined as $\sigma =  \frac{c_h}{H^* c_d}$
       \footnotemark[1]
        (where $c_h$ is the coefficient of heat and $H^*$ the enthalpy of vaporisation). 
        
        If the meteoroid survives this luminous trajectory or \textit{bright flight}, there is the possibility of recovering a meteorite on the ground. 
        Dedicated fireball camera networks such as the Desert Fireball Network (DFN) in Australia \citep{Bland2012} allow triangulated  trajectories of larger meteoroid bodies to be observed. Special shutters are used (in the case of the DFN, a liquid crystal shutter using modulated sequences \citep{Howie2016}) to encode timing throughout the trajectory. 
        Being able to predict the final state of the meteoroid is paramount to determining if there is any recoverable material, and is a necessary input to so-called \textit{dark flight} modelling (the process by which data from the luminous trajectory is converted into a fall line on the ground using atmospheric wind models), enabling likely search areas to be defined \citep{Ceplecha1987}. Accurately calculating a trajectory also allows the orbit for that body to be determined. Meteorites with orbits are rare; less than 0.05 \% of all meteorites. Knowing a meteorite's pre-atmospheric orbit gives contextual information to the picture they provide on early Solar System formation. Over time, the statistical analysis of calculated orbits may also assist in planetary defence of asteroidal debris streams.  

		Determining the state of a physical system based on a set of noisy measurements is known as filtering. The \emph{state} describes what a system is 'doing' at any given time. The flight path of an aircraft for example may be represented by its position, velocity and heading; position observations can be made in real time to estimate the velocity and heading of the aircraft. 
		Bayesian state-space estimation methods, such as the Kalman filter and its variants, address the filtering problem with the aim of estimating the true state of a system. 
		The adaptive approach predicts future states through a model of system equations and updates with respect to an observation. 
		Links between state variables defined in model equations allows unobserved state values to also be updated.
		
		This stochastic filtering approach suits the modelling of meteoroid trajectories using noisy and uncertain measurements. Typical meteoroid models mostly rely on measurements of the meteor/fireball brightness \citep{Kikwaya2011, Murray2000, Ceplecha2005}, though light curves tend to be variable and do not represent typical values predicted by single-body ablation models \citep{Campbell-Brown2004}.
        The meteoroid problem is complicated not only by unpredictable gross fragmentation in the atmosphere, but the majority of initial state parameters are entirely unknown ($m_0$, $\sigma$, $\kappa$).
        Multiple approaches have been taken to handle these unknowns in fireball trajectory analysis.        
        The manually intensive method of \citet{Revelle2007} is based on the brute force least squares approach of \citet{Ceplecha2005}. It does include the luminosity of the fireball (derived from manual interpretation of a light curve) as a proxy for mass loss and solves for fragmentation as well as $\sigma$ and $\kappa$. As it is still  based on a least squares optimisation, model and observation errors are not rigorously examined, rather overall errors are given as the standard deviation of residuals. 
        The amount of manual input required also limits the number of fireballs that may be analysed. The DFN observed over 300 fireball events in 2015 over its 2.5 million $km^2$ double station viewing area. This continental scale deployment of $>50$ automated observatories has been possible by the low cost of each system. At this time, there is no expensive, high voltage photomultiplier tube to measure fireball brightnesses.  A trajectory analysis approach that is able to determine meteoroid parameters without a light curve, and which can be automated, will allow an unbiased evaluation of all events.
        
        Very few models exist that enable the reduction of fireball data without a light curve.
        The method of \citet{Gritsevich2009} solves for two dimensionless parameters rather than multiple unknown trajectory parameters. This still requires an initial accurate velocity and struggles with highly scattered datasets \citep{Sansom2015}. 
        The various Kalman filtering methods used by \citet{Sansom2015} and \citet{Sansom2016} are fully automated techniques of determining the statistical likelihood of meteoroid state throughout bright flight, and allow a robust analysis of observation and model errors. 
	     As with previous dynamical approaches to fireball modelling, these require a pre-determined initial parameter set, withholding a general solution. To remove this limitation and fully analyse the statistical likelihood of the final state of a meteoroid given a range of likely initial states, we can use a method that combines a Monte Carlo (MC) approach to the filtering problem- a particle filter \citep{Gordon1993}. 
        Simply, a 'cloud' of particles are initiated with state values determined by a probability function. The 'cloud' will be denser where probabilities are higher. Particles are propagated forward in time according to the state equations and weighted according to an observation. A new generation of particles are \emph{resampled} from the existing pool, based on their weighting, and particles that are of low probability are preferentially removed. 
        
        The Bunburra Rockhole fireball was observed over the Australian outback by the DFN in 2007, and produced the network's first recovered meteorite \citep{Spurny2012}. An extended Kalman filter \citep{Sansom2015} and an unscented Kalman filter \citep{Sansom2016} have been used to model the Bunburra Rockhole fireball given a set of starting parameters. Neither filters explicitly include gross fragmentation, however \citet{Sansom2016} applied two Unscented Kalman filters in an Interactive Multiple Model to determine likely periods of fragmentation.
        Here we will examine the suitability of this sequential Monte Carlo technique for modelling fireball meteoroid trajectories using the Bunburra Rockhole fireball dataset.

    \section{Bayesian State-Space Estimation}
    \label{sec:bayes}

        The technique used in this paper for estimating meteoroid parameters is one of a broader class of techniques known as Bayesian state-space methods.  These methods involve encapsulating the knowledge of a system based on its \emph{state}, given by the vector $\mathbf{x}$.
        The state of an object could be its position and velocity, for example.  The probability of the object being in state $\mathbf{x}$ at time instant $t_k$ is represented as the conditional probability density function
        \begin{equation}
        \label{eq:1}
            p(\mathbf{x}_k|\mathbf{z}_{1:k}),
        \end{equation}
        where $\mathbf{z}_{k}$ is the observation of the system made at time $t_k$ and $\mathbf{z}_{1:k}$ is the history of all observations up until
        time $t_k$.  

        The calculation of \eqref{eq:1} is achieved recursively through the application of Bayes' rule
        \begin{equation}
            \label{eq:3}
            p(\mathbf{x}_k|\mathbf{z}_{1:k}) = \frac{p(\mathbf{z}_k|\mathbf{x}_{k})p(\mathbf{x}_k|\mathbf{z}_{1:k-1})}{p(\mathbf{z}_k|\mathbf{z}_{1:k-1})}.
        \end{equation}
        The terms in the numerator of (\ref{eq:3}) are defined through the \emph{state-space equations}, while the denominator can simply be considered as a normalising constant.

        There are three state-space equations.  The state \emph{prior} initialises the recursion and encapsulates all prior information about the state of the system
        \begin{equation}
            \label{eq:4}
            p(\mathbf{x}_0).
        \end{equation}
        The \emph{measurement equation} relates the observations (e.g. position) to the state of the system (e.g. position and velocity)
        \begin{equation}
            \label{eq:5}
            \mathbf{z}_k = h(\mathbf{x}_k,\mathbf{w}_k),
        \end{equation}
        where $\mathbf{w}_k$ is a stochastic noise process with known distribution.  Equation (\ref{eq:5}) defines the \emph{likelihood function}, $p(\mathbf{z}_k|\mathbf{x}_{k})$, which is the first term in the numerator of (\ref{eq:3}).  The \emph{process equation} models how the state evolves in discrete time
        \begin{equation}
            \label{eq:6}
            \mathbf{x}_{k+1} = f(\mathbf{x}_{k},\mathbf{u}_k),  
        \end{equation}
        where $\mathbf{u}_k$ is another noise process with known distribution. 
        Equation (\ref{eq:6}) defines the transition density $p(\mathbf{x}_{k+1}|\mathbf{x}_k)$, which is incorporated into the    second term in the numerator of (\ref{eq:3}) through the Chapman-Kolmogorov equation \citep{Jazwinski1970}
        \begin{equation}
            \label{eq:2}
            p(\mathbf{x}_k|\mathbf{z}_{1:k-1}) = \int p(\mathbf{x}_{k}|\mathbf{x}_{k-1}) p(\mathbf{x}_{k-1}|\mathbf{z}_{1:k-1}) d\mathbf{x}_{k-1}.
        \end{equation}

    \section{Meteoroid State-Space Equations}
    \label{sec:meter-state-estim}
        This section outlines the state-space and the state-space equations
        chosen to model the motion and measurement of a meteoroid process for
        the purposes of this paper.  The specific parameters used in the model
        to estimate the trajectory characteristics of the Bunburra Rockhole data-set are given in Section~\ref{sec:using-part-filt}.

        The state that defines the meteoroid system includes the physical parameters of motion, as well as trajectory parameters $\sigma$ and $\kappa$;
        \begin{equation}\label{eqn:state}
            \mathbf{x}_k =  \left[ \begin{array}{c}
            l_k \\
            v_k \\
            m_k \\
            \sigma_k \\
            \kappa_k
            \end{array} \right] 
            \begin{array}{l}
            \text{position} \\
            \text{velocity} \\
            \text{mass}  \\
            \text{ablation parameter} \\
            \text{shape-density parameter,}
            \end{array}
        \end{equation} 
        where the position is measured along a pre-defined path produced by triangulating observations from several imaging sensors.

        The measurement equation (\ref{eq:5}) is given by 
        \begin{align}\label{eq:9}
            \mathbf{z}_k = H\mathbf{x}_k +  \mathbf{w}_k,
        \end{align}
        where the measurement matrix is
        \begin{equation}
            \label{eq:7}
            H =  \left[ \begin{array}{ccccc}
            1 & 0 &  0 & 0 & 0
            \end{array} \right]
        \end{equation}
        and the measurement noise process, $\mathbf{w}_k$, is Gaussian with zero mean
        and variance $R_k$.

        As a meteoroid passes through the atmosphere, its behaviour can be modelled by the aerodynamic equations from the single body theory of meteoroid entry \citep{Hoppe1937, Baldwin1971} \eqref{eqn:dynamic}, which uses atmospheric densities, $\rho_a$, acquired using the NRLMSISE-00 atmospheric model \citep{Picone2002}, local
        acceleration due to gravity, $g$, and entry angle from horizontal, $\gamma_e$.  It is natural to model the change of meteoroid state as a continuous-time differential equation
        \begin{equation}
            \label{eq:8}
            \dot{\mathbf{x}} = f_c(\mathbf{x}) + \mathbf{u}_c,
        \end{equation}
        where $f_c(\mathbf{x})$ is defined using
        \begin{subequations} 
            \label{eqn:dynamic}
            \begin{align}
                \frac{dl}{dt} &    =v  \label{eqn:dyn_a}&& \\
                \frac{dv}{dt} &= -\frac{1}{2}\frac{c_d A \rho_a}{\rho_m^{2/3}} v^2 m^{(\mu -1)} +g  \sin{\gamma_e} &&=  - \kappa \rho_a   v ^2 m ^{(\mu -1)} + g  \sin{\gamma_e} \label{eqn:dyn_b}\\
                \frac{dm}{dt} &    =  -\frac{1}{2}\frac{c_h A \rho_a}{H^*} v^3 m^{\mu} &&=  -\sigma \kappa \rho_a v ^3 m ^\mu        \label{eqn:dyn_c}                       \\
                \frac{d\sigma}{dt} &    = 0  &&    \\
                \frac{d\kappa}{dt} &    = 0,  &&                                  
            \end{align}
        \end{subequations}
        and the continuous-time process noise, $\mathbf{u}_c$, is Gaussian
        with zero mean and covariance $\mathbf{Q}_c$.  Time integration of
        (\ref{eq:8}) is needed to arrive at the form required by the filtering
        state-space equation (\ref{eq:6}).  In this case
        \begin{equation}
            \label{eq:11}
            \mathbf{x}_{k+1} =  \int_{t_k}^{t_{k+1}}f_c(\mathbf{x}) \mathop{dt} + \mathbf{u}_k.
        \end{equation}
        Due to the non-linearities of (\ref{eqn:dynamic}) the discrete-time
        process noise, $\mathbf{u}_k$, is not Gaussian, but can be closely
        approximated by Gaussian noise with zero mean and covariance
        \begin{equation}
            \label{eq:Q}
            \mathbf{Q}_k = \int_{t_k}^{t_{k+1}}e^{Ft} \, \mathbf{Q}_c \, e^{F^T t} \mathop{dt}
        \end{equation}
        \citep{Grewal1993}, where the matrix $F$ is the linearised form of the process equation 
        \begin{equation}
            \label{eq:10}
            F = \frac{\partial  f_c(\mathbf{x})}{\partial \mathbf{x}}.
        \end{equation}
        Due to the form of the nonlinear functions (\ref{eqn:dynamic}), the
        integrations required by (\ref{eq:11}) and (\ref{eq:Q}) cannot be
        found analytically.  Numerical methods are used to calculate the
        integrals.

    \section{Particle Filter}
    \label{sec:particle-filter}

        There are a range of methods for finding the distribution of $\mathbf{x}_k$ by solving (\ref{eq:3}). The applicability of the method depends on the form of the state-space    equations.  If the measurement function and process function are linear and all the noise and prior distributions are Gaussian, then    the solution to (\ref{eq:3}) can be found analytically. This solution is known as the Kalman Filter \citep{Grewal1993}.  In the case where the equations are non-linear or the distributions are non-Gaussian, such as the single body equations for modelling meteoroid trajectory \eqref{eqn:dynamic}, there are no exact solutions and approximations are required.

        The Extended Kalman Filter (EKF) \citep{Sansom2015} approximates the noise distributions as Gaussian and finds a linear approximation to the process equations.  The Unscented Kalman Filter \citep{Sansom2016} approximates the posterior distribution as a Gaussian, but avoids approximating the measurement or process equations through a method of statistical linearisation \citep{Sarkka2007}.  

        A particle filter does not require any assumptions about the form of the state equations or have any limitations on the noise distributions.  This flexibility is achieved by representing the posterior density \eqref{eq:3} as a set of $N_s$ weighted particles, which are simply points in the state space \citep{Gordon1993,Arulampalam2002}.  The $i$th random particle at time $t_k$ is represented by its state, $\mathbf{x}_{k}^i$, and weight, $w_k^i$
        \begin{equation}
            \{\mathbf{x}_{k}^i, w_k^i \} \quad i = 1, ..., N_s .
        \end{equation}
        Weights are normalised so that 
        \begin{equation}\label{eqn:sum1}
            \sum_i^{N_s}  w_k^i = 1.
        \end{equation}
        The probability distribution of the state is approximated by this set of weighted particles
        \begin{equation}
            p(\mathbf{x}_k|\mathbf{z}_{1:k}) \approx \sum_i^{N_s} \delta(\mathbf{x}_k - \mathbf{x}_k^i) w_k^i,
        \end{equation}
        where $\delta(\mathbf{y})$ is the Dirac delta function, defined such that
        \begin{equation}
            \label{eq:13}
            \delta(\mathbf{y}) =
            \begin{cases}
                1 & \mathbf{y} = 0 \\
                0 & \text{otherwise}.
            \end{cases}
        \end{equation}
        Statistics can be computed on this set of particles, for example the
        mean of the distribution at any time $t_k$ is approximated by
        \begin{equation}\label{eqn:mean}
            \hat{\mathbf{x}}_k= \sum_i^{N_s} w_k^i\mathbf{x}_k^i,
        \end{equation}
        with the state covariance calculated as 
        \begin{equation}\label{eqn:var}
         Cov(\mathbf{x}_k) = \sum_i^{N_s}w_k^i (\mathbf{x}_k^i - \hat{\mathbf{x}}_k)(\mathbf{x}_k^i - \hat{\mathbf{x}}_k)^T. 
        \end{equation}
 
        There are strong similarities between the implementation of a particle filter and the simpler Kalman filter.  Both follow the three steps
        \begin{enumerate}
            \item Initialisation: start the filter with a known prior distribution, $p(\mathbf{x}_0)$
            \item Prediction: propagate the distribution from time $k-1$ to time $k$ using the process equation (\ref{eq:6})
            \item Update: use the measurement equation (\ref{eq:5}) to update the predicted distribution with the measurement information, producing the posterior distribution at time $k$, $p(\mathbf{x}_k|\mathbf{z}_{1:k})$
        \end{enumerate}
        The Kalman filter achieves these steps by exact analytic equations which manipulate the mean and covariance of the distribution at each step.  On the other hand the particle filter proceeds through calculation on each of the particles individually. 
        
        To initialise the particle filter, a set of particles are randomly sampled from the prior distribution, $p(\mathbf{x}_0)$, and weighted equally as $w_0^i = \frac{1}{N_s}$. 

        In the prediction step each particle is propagated forward in time via the process equation \eqref{eq:11}.  To incorporate the uncertainty of the system, a sample from the process noise, $u_k$, is randomly generated for each particle.  Using the process equation to propagate the particles results in the simplest form of the filter. The particle filter literature generalises this through importance sampling, where an arbitrary proposal distribution can be used, instead of the process equation \citep{Arulampalam2002}.  Sophisticated proposal distributions can make a particle filter implementation more efficient (require fewer particles), but they have not been investigated for this application. 
        
        The update step adjusts the weight of each particle.  The weight is obtained by evaluating the likelihood function for each particle
        \begin{equation}\label{eqn:obs}
            \tilde{w}_{k}^i  = p(\mathbf{z}_k|\mathbf{x}_{k}^i) w_{k-1}^i.
        \end{equation}
        The weights are then normalised to satisfy (\ref{eqn:sum1})
        \begin{equation}
            \label{eq:12}
            w_{k}^i = \frac{\tilde{w}_{k}^i}{\sum_i^{N_s} \tilde{w}_{k}^i}.
        \end{equation}
        Over time the particle weights can transfer to a few select particles, thereby updating insignificant particles at the expense of computing power \citep{Arulampalam2002}. This is known as the degeneracy problem and equation \eqref{eqn:neff} gives an approximate measure of particle effectiveness that can be used to assess the severity of the issue \citep{Arulampalam2002}.       
        \begin{equation}\label{eqn:neff}
            \hat{N}_\mathrm{eff} = \left(\sum_i^{N_s}(w_k^i)^2\right)^{-1}
        \end{equation}
        The degeneracy problem can be addressed by resampling the data after weights have been calculated. A new population of particles are generated from the current sample pool based on given weightings; the objective being to preferentially remove samples of lower weights. The probability of resampling any given particle $i$ is $w_k^i$. The optional resampling step is taken if the number of effective particles drops below some threshold.  After resampling all of the particle weights are set to $1/N_s$.

    \section{particle filter parameters for a meteoroid trajectory}
    \label{sec:using-part-filt}       
        Dedicated fireball networks, such as the DFN, capture fireball events from multiple locations, providing triangulated position observations with time.
        This also enables a rough calculation of velocities throughout the trajectory. 
        
        \subsection{Initialisation}
        When initialising the state prior for the set of $N_s$ particles at the start of the luminous trajectory ($t_0$), the initial position and, to an extent, the initial velocity\footnote{Determining $v_{inf}$ - or the velocity with which a body entered the Earth's atmosphere, as opposed to the 'initial' velocity that it has when its luminous trajectory is first observed, can be determined using reverse integration methods from the start of the luminous trajectory back to beyond the Earth's sphere of influence (e.g. \citet{Trigo2015}). This is done by the DFN data reduction process as part of orbital modelling. For the larger objects that generate fireballs (and that are the focus of this work) the difference between $v_{inf}$ and  $v_0$ is likely to be small, however a detailed discussion is outside the scope of this paper as the method described in this work (in accordance with others in the literature) model meteoroid bright flight only.} can be reasonably well constrained. 
        The other state parameters, $m, \sigma, \kappa$, however are not directly observable. 
        To explore the data space and determine likely values for $m_0$, as well as constants $\sigma$ and $\kappa$, each particle is initiated with a random value within a given range.
        The state prior for each particle is initialised according to Table \ref{table:init}, with $m_0^{min}$ in all cases set to 0.5 kg.

        \begin{table}[th!]
            \begin{center}
            \caption{Describes the method used by the particle filter to initialise state parameters for each particle. A random selection is made for each value using either a Gaussian probability density function (PDF) (mean and standard deviation given), a uniform PDF within a given value range or a multi-modal distribution in the case of bulk density.}
            \label{table:init}
            \begin{tabular}{c|c}
                \textbf{parameter} &  \textbf{method used}\\
                \textbf{to be initiated} & \\
                \hline
                $l_0$ & random choice based on Gaussian $\mathcal{N}(0, 10\,  m)$\\
                & (from triangulation errors)\\
                \hline
                $v_0$ & random choice based on Gaussian $\mathcal{N}(v_0, 500 \, m\, s^{-1})$\\
                & (from triangulation errors)\\
                \hline
                $m_0$ & random choice from 0 to $m_0^{max}$ (kg)\\
                \hline
                $\sigma$ & random choice between 0.001 to 0.05 $s^2\,km^{-2}$ \\
                & (from \citet{Ceplecha1998} for asteroidal material)\\
                \hline
                $\kappa$ & \begin{tabular}{cp{10cm}}$c_d$ - & random choice based on Gaussian $\mathcal{N}(1.3, 0.3)$ \\
                	& (based on aerodynamic drag values from \citet{Zhdan2007})\\
                $A$ -        & random choice based on Gaussian $\mathcal{N}(1.4, 0.33) $ \\
                & (close to spherical values)\\ 
                $\rho_m$ -      & the PDF representing meteorite bulk densities is multi-modal. To fully represent this distribution, initialisation is performed in two stages. \\& First, a random choice of meteorite type is made based on recovered percentages  
                (80 \% chondrites, 11 \% achondrites, 2 \% stony-iron, 5 \% iron, 2\% cometary \citep{Grady2000}). \\ & Second, a random choice of bulk density is made 
                based on the Gaussian PDF representing chosen meteorite type; \\
                & \begin{tabular}{lp{7.5cm}}
                    chondrites - &$\mathcal{N}(2700, 420)$ (after \citet{Britt2003}); \\
                    achondrites - &$\mathcal{N}(3100, 133)$ (after \citet{Britt2003}); \\
                    stony-iron - &$\mathcal{N}(4500, 133)$ (after \citet{Britt2003}); \\
                    iron - &$\mathcal{N}(7500, 167)$ (after \citet{Consolmagno1998}) ;  \\
                    cometary - &$\mathcal{N}(850, 117)$ (after \citet{Weissman2008}).  \end{tabular}
                \end{tabular}

            \end{tabular}
            \end{center}
        \end{table}

	\subsection{Prediction}
        At every observation time, $t_k$, the state of each particle is evaluated using the system model \eqref{eq:8}.  $\mathbf{Q}_c$ values used here to represent the continuous process noise in the given model for meteoroid trajectories are given by \eqref{eqn:Qc}. 
        The diagonal elements of $\mathbf{Q}_c$ in \eqref{eqn:Qc} are the variance values for $dl/dt$, $dv/dt$, $dm/dt$, $d\sigma/dt$, $d\kappa/dt$ respectively. The uncertainty in position and velocity are introduced through noise in the acceleration model \ref{eqn:dyn_b}, and the variance for $dl/dt$ for this process model is therefore set to $0\,m\,s^{-1}$.
        The other model equations however are not able to represent the system in its entirety; complications, such as fragmentation, affect all other state process models. 
        At this stage, we assume that the shape density and ablation parameters will not change dramatically over the meteoroid flight and are attributed small process noise values. 
        There is a high uncertainty in the mass loss for the single-body ablation model \ref{eqn:dyn_c} and so a large range of masses are allowed to be explored by the particles. The process noise in mass is a multiple of the mass in order to keep it within a consistent order of magnitude.
        The discrete process noise, $\mathbf{Q}_k$, is calculated at every time step following  \eqref{eq:Q}.
        \begin{eqnarray}
            \mathbf{Q}_c = 
			\left[\begin{array}{ccccc}
            (0\,m\,s^{-1})^2 & 0 & 0 & 0 & 0  \\
            0 & (75\,m\,s^{-2})^2 & 0& 0 & 0  \\
            0 & 0 & (0.2\times m_k\,\,kg\,s^{-1})^2 & 0 & 0 \\
            0 & 0 & 0 & (10^{-4} \, s\,km^{-2})^2 & 0 \\
            0 & 0 & 0 & 0 & (10^{-5}(SI)\,s^{-1})^2 \end{array} \right] \label{eqn:Qc}
        \end{eqnarray}
        To improve compute time of this method, the non-linear integration \eqref{eq:11} of all $N_s$ particles, and their associated $\mathbf{Q}_k$, is performed simultaneously using parallel multiprocessing. 

	\subsection{Update}
	    The triangulated position of the meteoroid along the trajectory at time $k$ is the  observation measurement $\mathbf{z}_k$.
        The weight $(\tilde{w}_{k}^i)$ for each particle, $\mathbf{x}_k^i$ is calculated using a one dimension Gaussian probability distribution function
        \begin{equation}
        p(\mathbf{z}_k|\mathbf{x}_{k}^i) = \frac{1}{\sqrt{2R_k\pi}}e^{-\frac{(\mathbf{z}_k-\mathbf{Hx}_{k}^i)^2}{2R_k}} 
        \end{equation}
		 in \eqref{eqn:obs}, with the observation noise having a variance $R_k = \left(100\, m\right)^2$. This is based on errors in timing and triangulated position, reflecting the accuracy of the data set being used.
	
	In order to avoid degeneracy in the particle set, we have use the stratified resampling method described by \citet{Arulampalam2002} after each update step.
	
    \section{Using a particle filter to predict a meteoroid trajectory}

        The data acquired by \citet{Spurny2012} for the Bunburra Rockhole fireball is used to test the suitability of the particle filter in estimating the state of a meteoroid during atmospheric entry. 
        The Bunburra Rockhole dataset consists of 113 published observations of position with time along the trajectory. Note that no observation data were published between $t = 0.0\,s$ and $t=0.1899 \,s$  or from  $t=5.3165\,s$ to $t=5.4589\,s$. Our modelling will use times relative to $t_0 = 0.1899 \,s$ along the trajectory.
        A particle filter is run using set of 10,000 particles ($N_s = 10,000$). Particles are initiated according to Table \ref{table:init} with $m_0^{max}$ set to 2,000 kg. 

\begin{figure} [htb!]
        	\centering
            \includegraphics[scale=0.3]{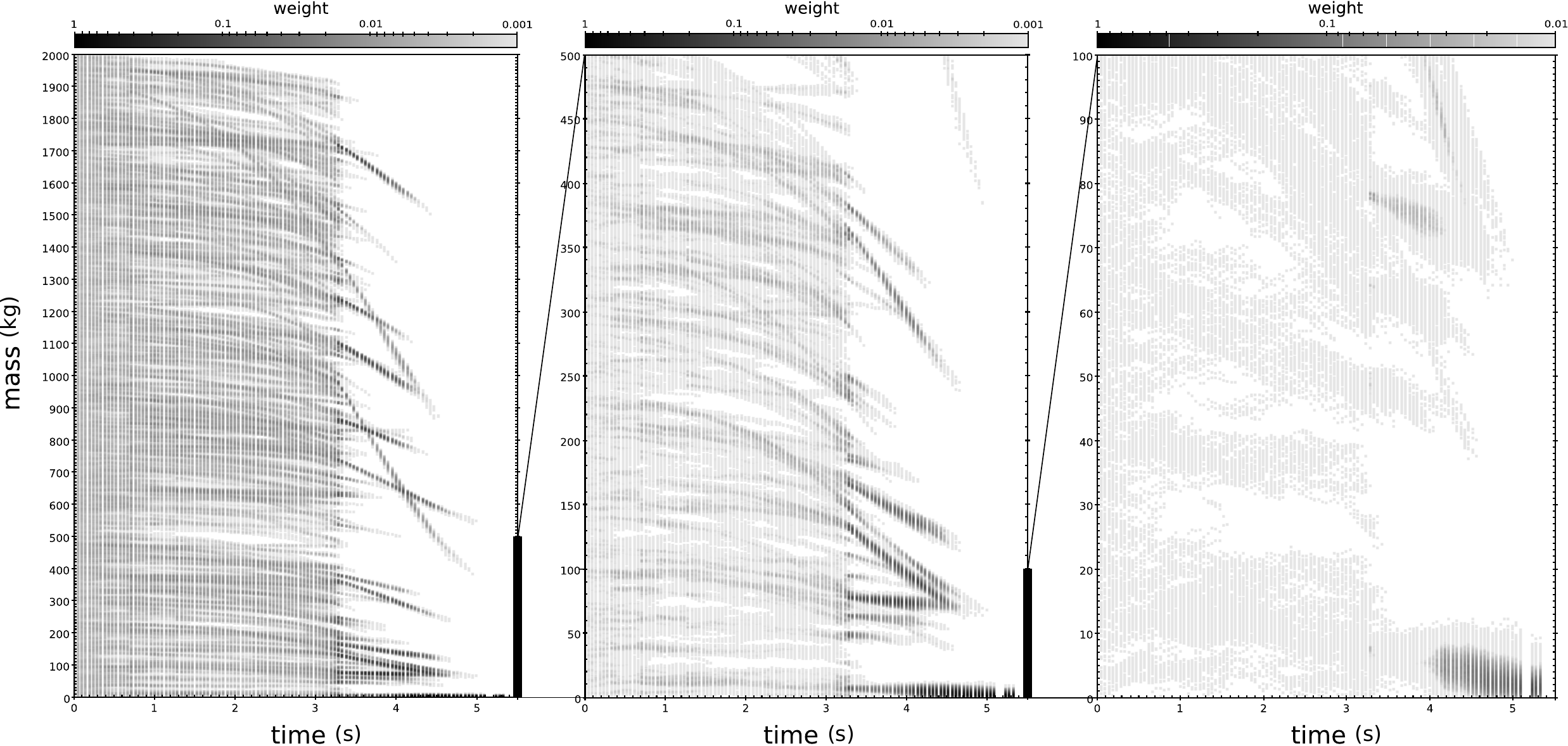}
            \caption{Mass estimates for particles, with  $w_k^i>0$, produced by the particle filter where $N_s = 10,000$, $m_0^{max}=2000\,kg$ were used and $\mathbf{Q}_c$ given by \eqref{eqn:Qc}. Colour scale is additive; weights of particles plotted in the same location are summed. Note the change in colour scale in the third frame to highlight $t_{end}$ weightings. At $t= 4.9\,s$ all particles with a weight greater than zero have a mass of 11\, kg or lower. Times correspond to the seconds since the 2nd recorded dash of the Bunburra Rockhole fireball; $t_0 = 0.1899\,s$ into the trajectory.  It is noticeable at $t_k=3.32\,s$ that there is a drastic reduction in the number of particle 'paths' that fit the observational data.  \label{fig:pf2000}}
        \end{figure}

        Figure \ref{fig:pf2000} shows all the resulting particle masses with weights $>0$ from $t_0$ to $t_{end}$. The range of $\sigma$ and $\kappa$ values used to initiate each particle results in a variety of predicted trajectory 'paths'.

\begin{figure}[th!]
        		\centering
            \includegraphics[scale=0.2]{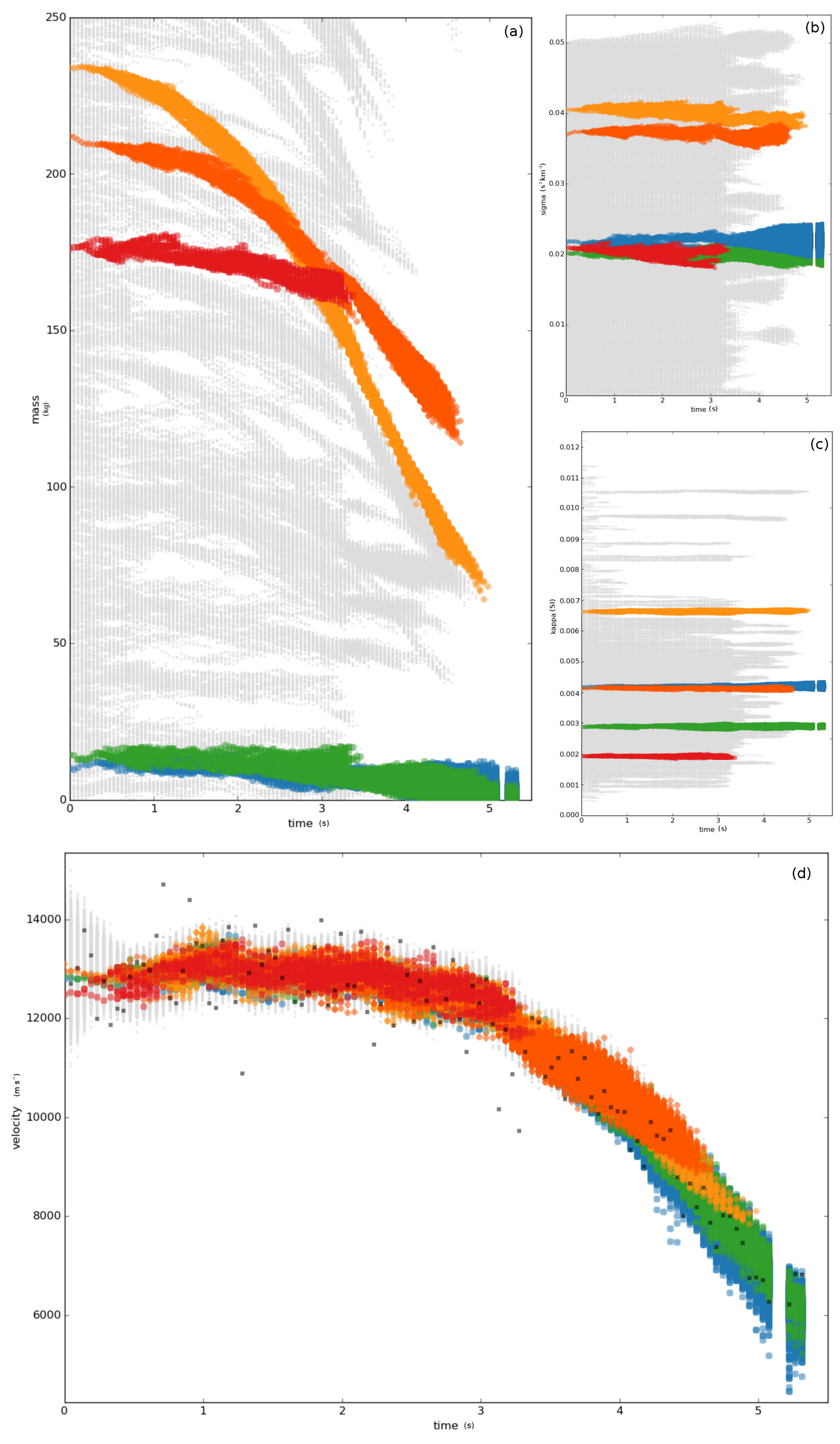}
            \caption{Particle states estimated by the particle filter. (a) Predicted mass with time. (b) Predicted ablation parameter,$\sigma$ with time. (c) Predicted shape density, $\kappa$, with time. (d) Predicted velocity with time. Particles originating from $\mathbf{x}_0^{a-e}$ (Table \ref{table:x0}) are highlighted with reference colours given in Table \ref{table:x0}). Note times correspond to seconds since the 2nd recorded dash of the Bunburra Rockhole fireball; $t_0 = 0.1899\,s$ into the trajectory.  It is noticeable at $t_k=3.32\,s$ that there is a drastic reduction in the number of particle 'paths' that fit the observational data. The parameter space after this time is much more constrained. \label{fig:pf250col}}
        \end{figure}

        To aid in understanding the different trajectories predicted by the particle filter, five particles at $t_0$ have been selected to discuss ($\mathbf{x}_0^j$ given in Table \ref{table:x0}). Figure \ref{fig:pf250col} highlights these particles, $\mathbf{x}_0^{a-e}$, along with all particles that are generated from them at later time steps 
        (either by propagation from $t_{k-1}$ or resampling at $t_k$). 
        
        \begin{table}[h]
            \begin{center}    
            \label{table:x0}
            \caption{The state of five particles at $t_0$ are shown. All future particles resampled from these are highlighted in Figure \ref{fig:pf250col} according to the colour given here.}
            \begin{tabular}{ccccccc}
                $\mathbf{ x}_0^j$& $l_0$ & $v_0$& $m_0$& $\sigma_0$& $\kappa_0$&  reference colour\\
                & $(m)$ & $(km\,s^{-1})$& $(kg)$& $(s^2\,km^{-2})$& (SI)& in Figure \ref{fig:pf250col}\\            
                \hline
                $\mathbf{ x}_0^a$ & $-1.57$ & $12.80$& $10.1$& $0.022$& $0.0083$& blue\\
                \hline
                $\mathbf{ x}_0^b$& $-18.60$ & $12.88$& $14.3$& $0.020$& $0.0058$& green\\
                \hline
                $\mathbf{x}_0^c$& $5.00$ & $12.48$& $176.2$& $0.021$& $0.0039$& red\\
                \hline
                $\mathbf{x}_0^d$& $-17.19$ & $12.96$& $212.1$& $0.037$& $0.0083$& dark orange\\
                \hline
                $\mathbf{x}_0^e$& $12.41$ & $13.10$& $234.0$& $0.041$& $0.0133$& light orange\\

            \end{tabular}
            \end{center}
        \end{table}

        The variation in $\sigma$ (Figure \ref{fig:pf250col}b) and $\kappa$ (Figure \ref{fig:pf250col}c) values with time is due to the addition of process noise, $\mathbf{u}_k$, in \eqref{eq:8}. As this noise is random Gaussian, it allows small variations between identical resampled particles that would have originally shared equal values. Areas of greater particle density are characteristic of higher probability states.

        Orange particles in Figure \ref{fig:pf250col} originate from $\mathbf{x}_0^e$. The steep change in mass with time (Figure \ref{fig:pf250col}(a) is due to the high $\sigma$ (Figure \ref{fig:pf250col}(b)) and $\kappa$ (Figure \ref{fig:pf250col}(c) values with which they were initiated. 
        Particles that no longer fit the observed data are preferentially removed by the resampling process and their 'path' discontinues in Figure \ref{fig:pf250col}. Although particles originating from $\mathbf{x}_0^{c-e}$ were initiated with diverse $\sigma$ (Figure \ref{fig:pf250col}b) and $\kappa$ (Figure \ref{fig:pf250col}c) values, they, along with all other particles with $m_0^i >27\,kg $ have insignificant weight past 5.0 seconds. A visual comparison of predicted particle velocities with velocities calculated from position measurements is shown in Figure \ref{fig:pf250col}d. The 'survival' of $\mathbf{x}_0^{a, b}$ to $t_{end}$ is due to their higher $w_k^i$ values indicating superior fits to the observations (and visually noticeable in Figure \ref{fig:pf250col}d).
        
		The final trajectory parameters of the Bunburra Rockhole meteoroid have been previously determined by \citet{Spurny2012} using the dynamic gross fragmentation model (GFM) of \citet{Ceplecha1993} and the meteoroid fragmentation model (MFM) of \citet{Ceplecha2005} which integrates fireball brightness with the dynamics (Table \ref{table:pf1_results}). Both the GFM and MFM require initial assumptions including the entry mass and a manually pre-defined fragmentation pattern based on the light curve \citep{Ceplecha2005}. Errors given by these models relate to the standard deviation of the residuals between modelled and measured observations; observational uncertainties, assumptions made in the model and model parameters are not propagated.
		The Kalman filter methods applied by \citet{Sansom2015} and \citet{Sansom2016} to meteoroid trajectory modelling perform a comprehensive analysis of the errors of both model and observations but share the limitations of previous models in requiring a single set of initial entry parameters to be pre-determined. 
		
		The statistical approach of the particle filter is not limited to any one set of input parameters. It encapsulates all prior knowledge of the parameter space by exploring the full range of plausible parameter values to produce an unbiased analysis.
		 Given that model and observation uncertainties are incorporated and propagated, this method provides a statistically robust final state estimate which is no longer dependent on any single set of assumed input parameters, providing a more realistic understanding of real-world variability. The independence of the particle filter and lack of manual input enables full automation of this method.		
		\begin{table}[ht!]
				\centering
			\caption{Mean final state values estimated by the particle filter \eqref{eqn:mean}, alongside published values. Errors given by all previous methods reflect only model errors within the given initial input assumptions given. The GFM and MFM methods do not consider observation uncertainties \citep{Ceplecha2005}. The particle filter  errors are calculated as $\sqrt{Var(\hat{\mathbf{x}}_k)}$ given by \eqref{eqn:var}, and alone gives a fully inclusive analysis of trajectory model and observation uncertainties to provide a more realistic understanding of real-world variability.\label{table:pf1_results}}
			\begin{tabular}{c|c|c|c|c|c}
				& $l_{end}$ & $v_{end}$ & $m_{end}$ & $\sigma_{end}$ & $\kappa_{end}$ \\
				& ($km$) & ($km\,s^{-1}$) & ($kg$) & ($s^2\,km^{-2}$) & ($SI$) \\
				\hline\\
				GFM\footnotemark[3] & & & $1.5\pm0.2$ & $0.0331\pm0.0007$ & \\
				&&&&(apparent)&\\
				MFM\footnotemark[3] & & $5.77\pm0.04$ & $1.1$ &  $0.002\pm0.001/0.004$  & $0.0035$\\
				&&&&(intrinsic)&\\
				Dynamic\footnotemark[4]  & $60.07$ & $6.109$ & $2.36$ & 0.0371 & $0.0062$\\
				optimisation&&&&(apparent) &\\
				EKF\footnotemark[4]  & $60.03\pm0.062$ & $6.05\pm0.24$ & $2.30\pm1.63$ &&\\
				&&&&&\\
				UKF\footnotemark[5]  & $60.04\pm0.058$ & $6.10\pm0.20$& $2.88\pm1.04$ & &\\
				&&&&&\\
				IMM\footnotemark[5]  & $60.01\pm0.007$  & $5.90\pm0.06$& $1.32\pm0.49$ & &\\
				&&&&&\\
				Particle filter & $59.89\pm0.038$ & $6.03\pm0.22$& $2.16\pm1.33$& $0.0219\pm0.0007$& $0.0042$\\
				&&&&(apparent)&$\pm0.000$\\
			\end{tabular}
		\end{table}
		\footnotetext[3]{ \citet{Spurny2012}; GFM = gross fragmentation model; MFM = meteoroid fragmentation model.}
		\footnotetext[4]{ \citet{Sansom2015}; $\kappa$ value determined using $c_d=1.3$; EKF = extended Kalman filter.}
		\footnotetext[5]{\citet{Sansom2016}; UKF = unscented Kalman filter; IMM = interactive multiple model.}
		
        The spread of final particle states at $t_{end}$ can be summarised by the weighted mean \eqref{eqn:mean} in Table \ref{table:pf1_results}. Errors are calculated as the square root of the covariance diagonal elements given by Equation \eqref{eqn:var}.
        The ablation parameter is an interesting result. 
        Although the particle filter does not explicitly model fragmentation, $Q_c$ allows for a certain amount of variation in state parameters due to un-modelled processes and inherently includes fragmentation to some extent, without the need for a pre-defined fragmentation pattern (required by MFM \citep{Ceplecha2005}). 
        As discussed by \citet{Ceplecha2005}, the intrinsic value of the ablation parameter remains constant throughout the trajectory regardless of fragmentation. When fragmentation is not modelled explicitly, variations in the ablation parameter appear to occur and must therefore be expressed as the \textit{apparent} ablation parameter. The GFM produces an \textit{apparent} $\sigma$ whereas the MFM, as it incorporates the light curve, is able to define the \textit{intrinsic} $\sigma$. The value determined using the particle filter is slightly lower than the apparent $\sigma$ of the GFM and it is therefore plausible that we can use this difference to quantify the extent to which fragmentation is included in the final state estimate.
        
        Using a particle filter the state estimates at each time step are iteratively updated based on the past data; future observations are not included. The final states alone result from processing all observations. As a predicted particle  becomes inconsistent with the observations, it becomes an unlikely scenario for future times but it does not mean this original path can be discounted. 
        It is noticeable at $t_k=3.32\,s$ that there is a drastic reduction in the number of particle 'paths' that fit the observational data. The parameter space after this time is much more constrained. 
        All particles at $t_{end}$ originate from particles with  $\mathbf{x}_0<27\,kg$;
        these particles are consistent with both parts of the trajectory displaying no dramatic change in mass. 
        It is possible that particles of initially higher mass are discontinued in favour of lower mass scenarios as a result of gross fragmentation reflected in the observation data.
        Without including all the data at every time-step, the most likely state 'path' for the entire trajectory cannot be constrained; we cannot distinguish the full particle history. 
        
        In order to distinguish likely initial masses, we need to be able to explore drastic changes in mass. The interactive multiple model (IMM) smoother as described by \citet{Sansom2016} has this capability and uses all observational data at each time step. It however requires a single pre-defined set of initial parameters. This is a well suited complementary method to our current implementation of a particle filter. 
        The particle filter framework however is flexible enough to incorporate dynamic models that explicitly capture gross fragmentation events.  Future work will explore more sophisticated dynamic models as well as particle filter smoothing to reconstruct the full meteoroid trajectory.
        
         Including brightness as a state in trajectory modelling would also provide an additional observation with which to weight particles. As brightness is linked to mass, its addition would not only improve state estimates, but would inherently include information on fragmentation. 

    \section{Conclusion}
    The use of a particle filter to approximate fireball trajectories provides a statistical analysis of the meteoroid state, including unobservable trajectory parameters. This is the first approach of its kind in this field. 
    Other non-linear filtering algorithms such as the Extended Kalman filter \citep{Sansom2015} and the Unscented Kalman filter \citep{Sansom2016}, as well as other least-squares approaches \citep{Ceplecha1993, Ceplecha2005}, require a pre-determined set of initial parameters to statistically analyse the trajectory of a meteoroid. 
    The iterative Monte Carlo simulations of a particle filter is not only capable of automating the analysis of fireball trajectories, but is able to do so without the need for limiting input parameters to single assumed values, rather it encapsulates all prior knowledge of the parameter space, to produce an unbiased analysis.
    The adaptive filter approach uses the observations of the meteoroid's position as it travels through the Earth's atmosphere to update state estimates. Predicted positions similar to those observed are given a higher weighting and are preferentially resampled at the next time step. This gives a final state estimate (Table \ref{table:pf1_results}) with robust error propagation of uncertainties in the initial parameters, observations and the dynamic model (e.g. unpredictable gross fragmentation events). 
    Even though trajectory parameters $\sigma$ and $\kappa$ are not currently set to vary systematically with time (noise is added to create diversity between resampled particles to avoid degeneracy only), a stochastic approach to their determination has not previously been conducted.
    Incorporating brightness as an additional state will provide supplementary data and improve estimates. This method currently allows an automated dynamic analysis of fireball trajectories. 
    
	\section*{Acknowledgements}
	This work was funded by the Australian Research Council as part of the Australian Laureate Fellowship scheme, and supported by resources provided by the Pawsey Supercomputing Centre with funding from the Australian Government and the Government of Western Australia. 

    \bibliography{library}{}
    \bibliographystyle{abbrvnat}

\end{document}